\documentclass[12pt]{iopart}
\usepackage{iopams,graphicx,cite}

\begin{document}

\title[Photoabsorption spectra of the diamagnetic hydrogen atom]
 {Photoabsorption spectra of the diamagnetic hydrogen atom in
 the transition regime to chaos: Closed orbit theory with bifurcating orbits}
\author{Toma\v z Fab\v ci\v c\dag, J\"org Main\dag, Thomas Bartsch\ddag\ and
G\"unter Wunner\dag}
\address{\dag\ Institut f\"ur Theoretische Physik 1, Universit\"at Stuttgart,
 70550 Stuttgart, Germany}
\address{\ddag\ Center for Nonlinear Science, School of Physics, Georgia 
 Institute of Technology, Atlanta, GA 30332-0430, USA}
\date{\today}

\begin{abstract}
With increasing energy the diamagnetic hydrogen atom undergoes
a transition from regular to chaotic classical dynamics, and the
closed orbits pass through various cascades of bifurcations.
Closed orbit theory allows for the semiclassical calculation of 
photoabsorption spectra of the diamagnetic hydrogen atom.
However, at the bifurcations the closed orbit contributions diverge.
The singularities can be removed with the help of uniform semiclassical
approximations which are constructed over a wide energy range for 
different types of codimension one and two catastrophes.
Using the uniform approximations and applying the high-resolution harmonic 
inversion method we calculate fully resolved semiclassical photoabsorption 
spectra, i.e., individual eigenenergies and transition matrix elements
at laboratory magnetic field strengths, and compare them with the results
of exact quantum calculations.
\end{abstract}


%\maketitle

\section{Introduction}
Rydberg atoms in a magnetic field have become  prototype examples
of a quantum system with an underlying classical dynamics changing
from regular to chaotic motion with increasing excitation energy
\cite{Fri89,Has89,Wat93}.
The Garton-Tomkins resonances originally found in barium atoms
\cite{Gar69} and similar types of experimentally observed long-range
modulations \cite{Mai86,Hol88,Mai91,Kar99} can be associated with 
classical closed orbits starting at and returning back to the nucleus.
A deeper quantitative analysis and interpretation of these features
is possible within semiclassical theories, such as periodic orbit theory
\cite{Gut90} and, as a variant for the photoabsorption of atomic systems,
closed orbit theory \cite{Du88,Bog89}.
In these theories, either the density of states or atomic photoabsorption 
spectra, as functions of energy, are given as superpositions of a smoothly 
varying part and sinusoidal modulations, whose frequencies, amplitudes, and 
phases are given in terms of the classical parameters of the orbits.

While closed orbit theory has been successfully applied to the 
interpretation of quantum spectra in terms of the closed orbits of the
underlying classical system \cite{Mai94,Cou95,Kip98,Kip99}, the inverse
procedure, i.e., the semiclassical calculation of the energies and transition
strengths of individual eigenstates, is much more challenging for the
following reasons:
Firstly, closed orbit theory formally requires the knowledge of the infinite 
set of all closed orbits, which is impossible to obtain for nonintegrable 
systems where the orbits must be searched numerically.
Secondly, in both periodic orbit and closed orbit theory the infinite sum
over all orbit contributions suffers from fundamental convergence problems,
and, thirdly, in generic systems the orbits undergo bifurcations when the
energy is varied, and the contributions of isolated orbits exhibit unphysical
singularities at these bifurcation points.

Substantial progress has already been achieved to overcome these problems
separately:
On the one hand, the bifurcations of closed and periodic orbits of the 
diamagnetic hydrogen atom have been investigated in \cite{Mao92,Sad95,Sad96}
and the divergences of isolated orbit contributions at bifurcations have 
been removed with the help of uniform semiclassical approximations for 
various types of bifurcations of codimension one and two 
\cite{Mai97a,Mai98a,Bar99a,Bar99b}.
On the other hand, the harmonic inversion technique based on high-resolution 
signal processing has been introduced as a method for semiclassical 
quantization of generic systems \cite{Mai97c,Mai98b,Mai99d}.
This method allows one to extract discrete eigenenergies and matrix elements 
from a finite set of classical orbits, and thereby circumvents the convergence
problems of the infinite sums in closed orbit or periodic orbit theory.
In its original form it was applied to spectra of the diamagnetic hydrogen 
atom at constant scaled energy \cite{Mai99a,Mai99b} where the need 
to account for the effects of bifurcations does not arise.
A semiclassical quantization with bifurcating orbits that merges these
independent strands of research has only recently been achieved for 
an integrable atomic system, viz.\ the hydrogen atom in an electric 
field \cite{Bar02,Bar03a}.
Energy dependent photoabsorption Stark spectra have been obtained by
considering the bifurcations of the ``uphill'' and ``downhill'' orbit 
parallel and antiparallel to the direction of the external electric field.

In this paper we demonstrate that the semiclassical quantization with 
bifurcating orbits can be successfully applied to a more challenging system, 
viz.\ the hydrogen atom in a magnetic field, where the classical equations 
of motion are nonseparable and the dynamics undergoes a transition from 
regular to chaotic dynamics with increasing energy.
The bifurcation scenarios encountered there are much more complicated 
than those in the hydrogen atom in an electric field, and different types 
of catastrophes with codimension one and two must be used to remove the 
divergences at the bifurcations.

Although the numerical effort for the calculation of the semiclassical 
photoabsorption spectrum of the diamagnetic hydrogen atom is much higher
than for the corresponding exact quantum computations, the results of this
paper are of fundamental interest for the development, understanding, and 
practical applications of semiclassical theories.
As Einstein \cite{Ein17} pointed out as early as 1917, the ``old'' quantum 
theory based on the Bohr-Sommerfeld quantization rules is doomed to
failure when applied to nonintegrable systems.
About ninety years later we have now succeeded in obtaining the 
high-resolution photoabsorption spectra of a nontrivial atomic system with 
mixed regular-chaotic dynamics semiclassically from first principles.
The necessary ingredients are closed orbit theory, uniform semiclassical 
approximations at bifurcations, and the harmonic inversion method.

The paper is organized as follows.
In Sec.~\ref{class_dyn:sec} the classical dynamics of the hydrogen atom
in a magnetic field and various types of closed orbit bifurcations are
discussed.
In Sec.~\ref{uni:sec} closed orbit theory is introduced and the uniform
approximations at bifurcations of closed orbits are constructed.
Semiclassical high-resolution photoabsorption spectra with individual
eigenenergies and transition matrix elements are obtained by
application of the harmonic inversion method in Sec.~\ref{high_res_spectra:sec}
and are compared with exact quantum spectra.
Concluding remarks are given in Sec.~\ref{conclusion:sec}.

\section{Classical dynamics and closed orbit bifurcations}
\label{class_dyn:sec}
The classical dynamics of the diamagnetic Kepler problem has 
already been discussed extensively in the literature 
(for reviews see, e.g., \cite{Fri89,Has89,Wat93}).
Here we briefly recapitulate the essentials which are necessary to 
what follows.
The Hamiltonian in atomic units 
(with the magnetic field ${\bf B}=B{\bf e}_z$ along the $z$-axis,
$\gamma\equiv B/(2.35\times 10^5\,{\rm T})$, and angular momentum component 
$L_z=0$) reads
\begin{equation}
\label{Ham1:eq}
 H = \frac{1}{2}{\bf p}^2 - \frac{1}{r} + \frac{1}{8}\gamma^2(x^2+y^2) = E \; ,
\end{equation}
where $E$ is the energy.
Using semiparabolic coordinates
$\mu=\sqrt{r+z}$,
$\nu=\sqrt{r-z}$
the Hamiltonian (\ref{Ham1:eq}) can be transformed to
\begin{equation}
\label{Ham2:eq}
 h = \frac{1}{2}(p_\mu^2+p_\nu^2) - E (\mu^2+\nu^2)
   + \frac{1}{8}\gamma^2\mu^2\nu^2(\mu^2+\nu^2) = 2 \; .
\end{equation}
Note that Hamilton's equations of motion derived from (\ref{Ham2:eq}) 
are free of singularities at the Coulomb centre.
The Hamiltonian~(\ref{Ham1:eq}) is invariant under a reflection in the
$xy$-plane perpendicular to the magnetic field. As a consequence, all
orbits that are not in that plane occur in symmetric pairs.
The closed orbits leave the nucleus ($r=0$) with an initial angle $\vartheta_i$
to the $z$-axis and return back to the origin with final angle $\vartheta_f$
after time period $T$.
The stability properties of the closed orbits are given in terms of the
$2 \times 2$ monodromy matrix $\mathbf M$, which linearly maps local deviations
($\delta q,\delta p$) of the starting point in the directions perpendicular to
the orbit in coordinate and momentum space onto local deviations of the final
point:
\begin{equation}
   \left(\matrix{\delta q(T) \cr \delta p(T)} \right)
 = {\mathbf M} 
   \left(\matrix{\delta q(0) \cr \delta p(0)} \right)
 = \left(\matrix{m_{11} & m_{12} \cr m_{21} & m_{22}}\right)
   \left(\matrix{\delta q(0) \cr \delta p(0)} \right) \; .
\label{mono:eq}
\end{equation}
Closed orbits bifurcate when the element $m_{12}$ of the monodromy matrix 
vanishes.
(Note that a different condition $\det({\mathbf M}-{\bf 1})=0$ is valid
for periodic orbit bifurcations.)

The classical dynamics does not depend on the energy $E$ and magnetic field
strength $\gamma$ separately.
Instead, the scaled Hamiltonian $\tilde H=H\gamma^{-2/3}$ is independent of 
$\gamma$ if it is expressed in terms of the scaled semiparabolic coordinates 
$\tilde\mu=\gamma^{1/3}\mu$ and $\tilde\nu=\gamma^{1/3}\nu$, so that the 
scaled energy $\tilde E=E\gamma^{-2/3}$ is the only control parameter.
The time and classical action scale as $\tilde t = t\gamma$ and 
$\tilde s = s\gamma^{1/3}$, respectively, and the matrix element $m_{12}$ 
of the monodromy matrix $\mathbf M$ in equation~(\ref{mono:eq}), which is 
important in closed orbit theory, scales as 
$\tilde m_{12} = m_{12}\gamma^{1/3}$.

In the limit of infinitely negative energy ($E\to -\infty$) only two 
closed orbits parallel and perpendicular to the magnetic field axis exist.
Their multiple repetitions are called basic vibrators $V_\mu$ and rotators 
$R_\mu$, respectively \cite{Hol88,Mai91}, where the index $\mu$ gives the 
number of repetitions.
When the energy is increased, the basic vibrators and rotators
undergo cascades of bifurcations where new closed orbits $V_\mu^\nu$
and $R_\mu^\nu$ are created in a systematic way 
(see Secs.~\ref{V_bif:sec} and \ref{R_bif:sec} below).
These orbits can run through further bifurcations as discussed in
Sec.~\ref{V_R_bif:sec}.
Furthermore, new closed orbits which are not directly related to the
bifurcation tree of the basic vibrators and rotators are created 
``out of nowhere'' by tangent bifurcations.
With increasing energy a transition from nearly regular to chaotic phase
space takes place, along with a rapid proliferation of closed orbits, and
thus the semiclassical quantization with bifurcating orbits becomes more 
and more challenging with growing energy.

\subsection{Bifurcations of the parallel orbit}
\label{V_bif:sec}
With increasing energy each repetition $V_\mu$ of the parallel orbit 
undergoes an infinite sequence of bifurcations with an accumulation point 
at the field free ionization threshold $E=0$.
Individual bifurcations are counted by integer numbers $\nu$.
\begin{figure}
\begin{center}
\includegraphics[width=0.85\columnwidth]{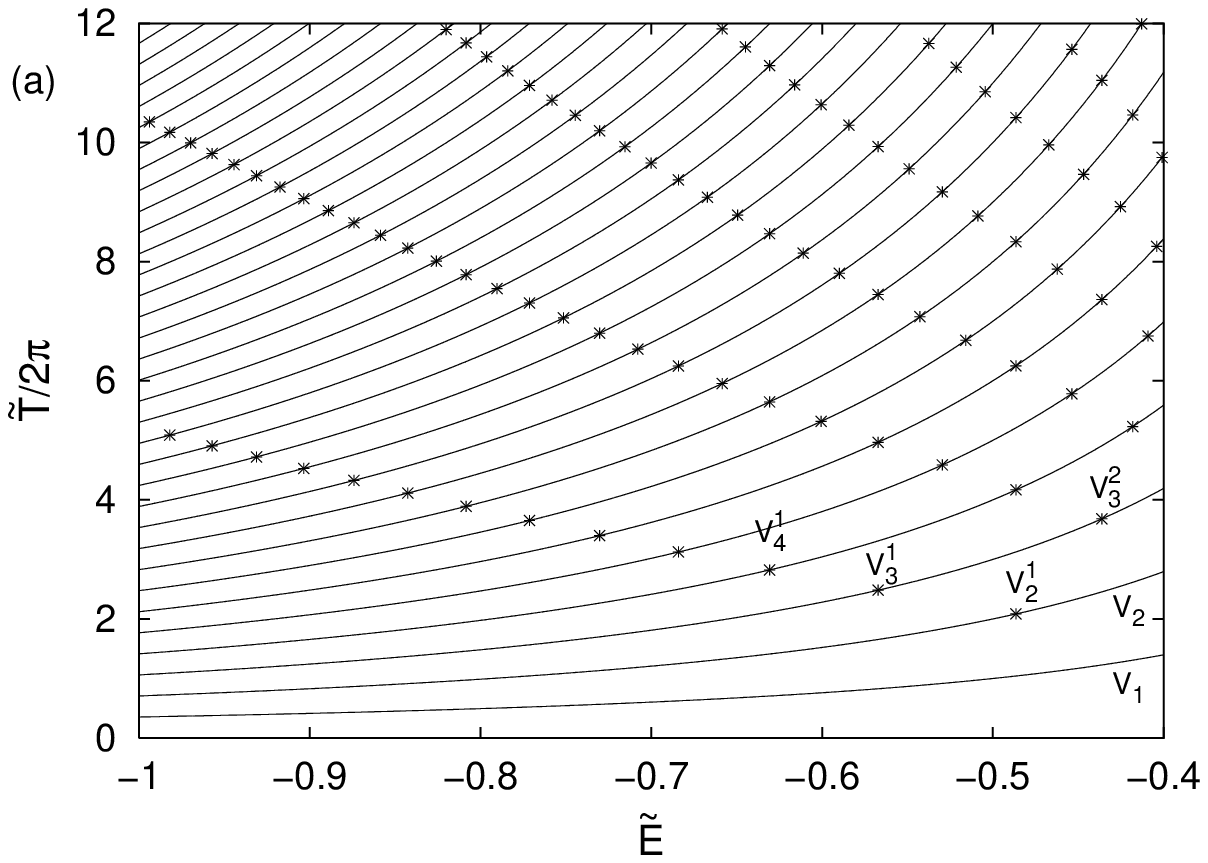}
\includegraphics[width=0.85\columnwidth]{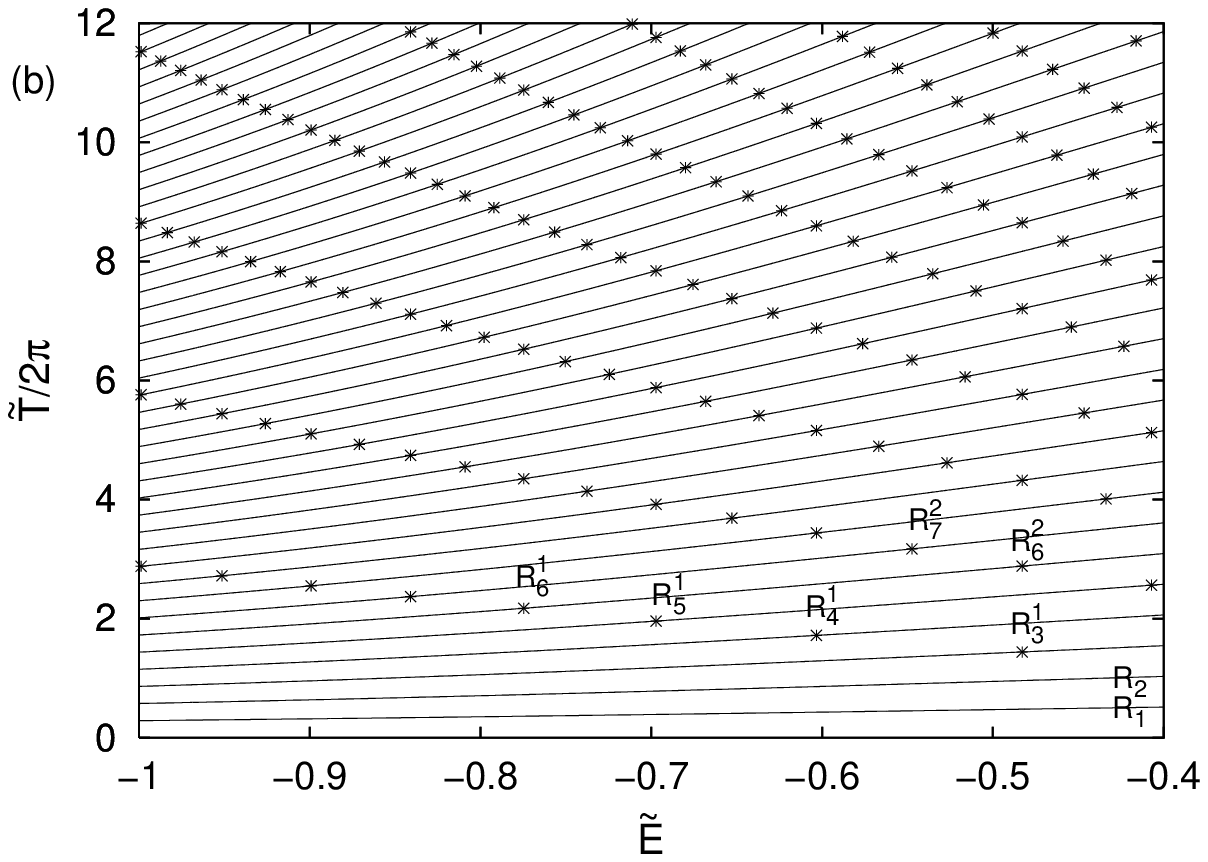}
\end{center}
\caption{Periods of (a) the basic vibrators $V_\mu$ and (b) the basic 
rotators $R_\mu$.  The bifurcations are marked by crosses, some of them 
are labelled by the symbols $V_\mu^\nu$ or $R_\mu^\nu$ of the bifurcating 
closed orbits.  The periods of the bifurcating orbits are not shown to 
keep the figure concise.}
\label{fig1}
\end{figure}  
The bifurcations of the parallel orbit are illustrated in figure~\ref{fig1}a,
where the periods of the basic vibrators $V_\mu$ are plotted as functions 
of the scaled energy $\tilde E$.
At the bifurcation points new orbits called $V_\mu^\nu$ \cite{Hol88,Mai91} 
are created.
The bifurcation points are marked by crosses in figure~\ref{fig1}a, and some 
of them are labelled by the symbols $V_\mu^\nu$ of the bifurcating closed 
orbits.
The periods of the bifurcating orbits $V_\mu^\nu$ as functions of the
scaled energy are not shown to keep the figure concise.
From figure~\ref{fig1} it becomes evident that bifurcations occur rather 
frequently in energy, and the correct handling of bifurcations is of
crucial importance for the semiclassical quantization of this system.

\begin{figure}
\begin{center}
\includegraphics[width=0.8\columnwidth]{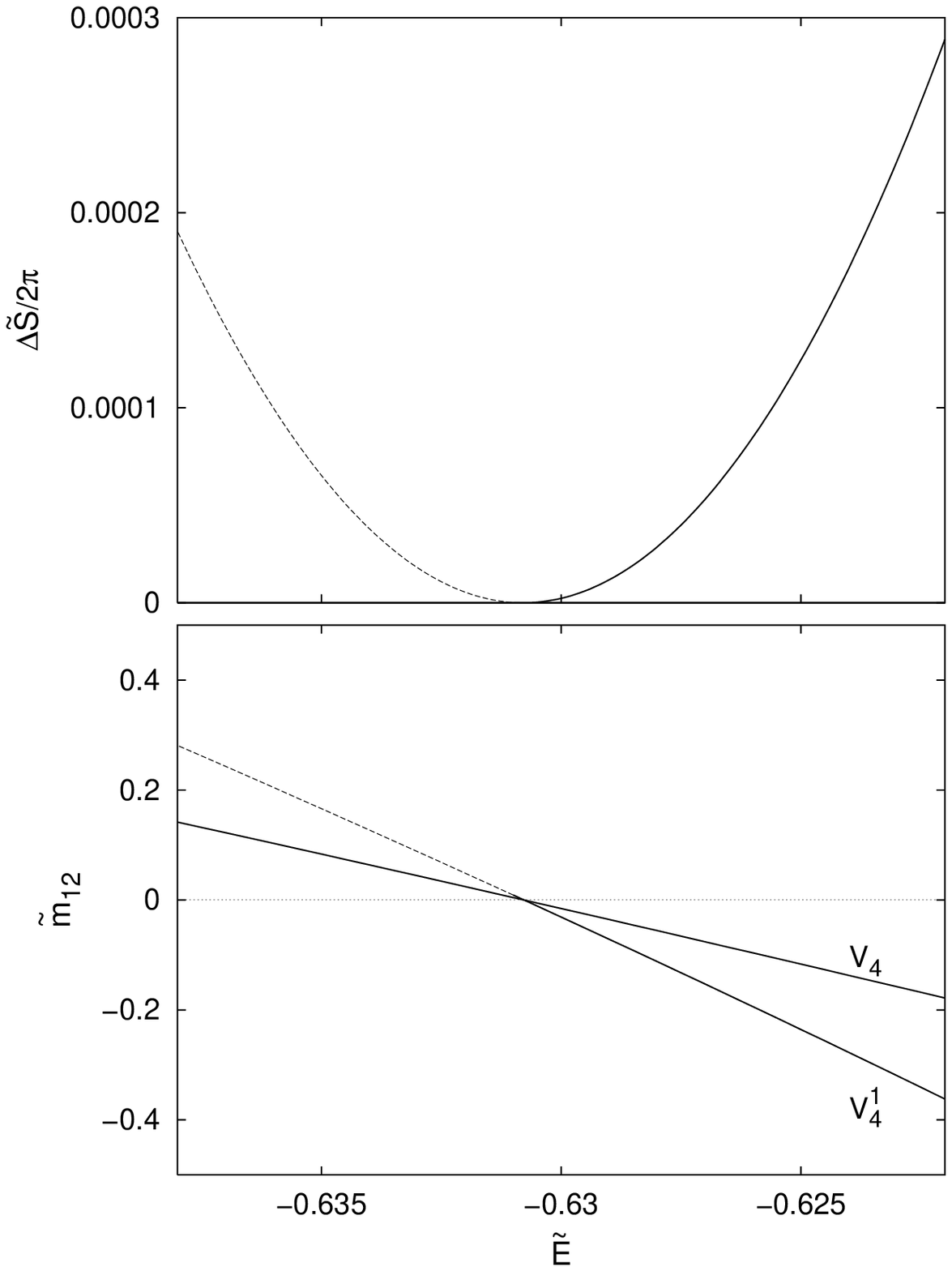}
\end{center}
\caption{Closed orbit parameters for the pitchfork bifurcation of the
orbits $V_4$ and $V_4^1$.
Solid and dashed lines mark the real and ghost orbits, respectively.}
\label{fig2}
\end{figure}

The bifurcations of the parallel orbit of the diamagnetic hydrogen
atom resemble those of the ``uphill'' and ``downhill'' orbit of the 
hydrogen atom in an electric field \cite{Gao94,Bar02,Bar03a}.
The system is rotationally symmetric around the parallel (axial) orbit,
and a bundle of three-dimensional non-axial orbits
collide with the axial orbit in a pitchfork bifurcation.
The non-axial orbits are real orbits or complex ``ghost'' orbits in the 
complex continuation of the phase space when the value of $E$ is above or 
below the bifurcation energy, respectively.
As an example, the bifurcation scenario of the orbits $V_4$ and $V_4^1$ 
is illustrated in figure~\ref{fig2}.
The difference $\Delta\tilde S/2\pi=(\tilde S_{\rm ax}-\tilde S_{\rm
non})/2\pi$ between the scaled actions of the axial (ax) and the non-axial
(non) orbit and the element $\tilde m_{12}$ of the scaled monodromy matrix
are shown as functions of the scaled energy.  The solid and dashed lines
refer to the real and ghost orbits, respectively.  The approximately linear
dependence of $\tilde m_{12}$ on the energy and the quadratic behaviour of
$\Delta\tilde S$ around the bifurcation point is typical of pitchfork
bifurcations.  For the ghost orbits the initial angle $\vartheta_i$ (not
shown in figure~\ref{fig2}) is purely imaginary.  The classical action and
monodromy matrix, however, are real valued.  Note that the closed orbit
parameters of both the real and ghost orbits are required for the
construction of the uniform semiclassical approximations in
Sec.~\ref{uni:sec}.

\subsection{Bifurcations of the perpendicular orbit}
\label{R_bif:sec}
The multiple repetitions of the orbit perpendicular to the magnetic field,
i.e., the basic rotators $R_\mu$ also undergo a sequence of bifurcations,
where new closed orbits $R_\mu^\nu$ are created \cite{Hol88,Mai91}.
Contrary to the bifurcations of the basic vibrators, the number of 
bifurcations of a rotator is finite.
The integer $\nu$ that identifies the bifurcations is limited by $\nu < \mu$.
An overview of the bifurcations of the perpendicular orbit is given in
figure~\ref{fig1}b, where the bifurcation points are marked by crosses, 
with some of them labelled by the symbols $R_\mu^\nu$.

\begin{figure}
\begin{center}
\includegraphics[width=\columnwidth]{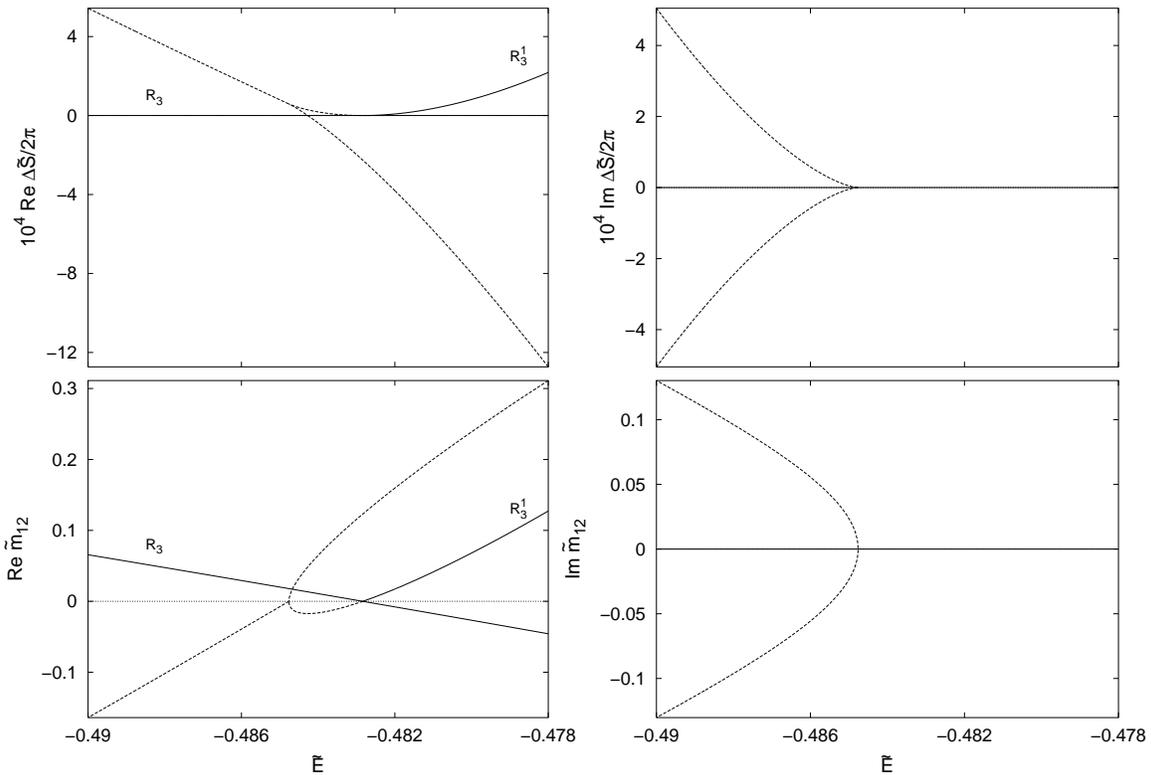}
\end{center}
\caption{Closed orbit parameters for the bifurcation scenario where the
orbit $R_3^1$ is created from $R_3$.  A sequence of a tangent bifurcation
where only ghost orbits participate and a nearby pitchfork bifurcation is
the typical scenario for all bifurcations of the perpendicular orbit at
scaled energies $\tilde E < -0.418$.  Real and ghost orbits are marked by
solid and dashed lines, respectively.}
\label{fig3}
\end{figure}

The bifurcations of the basic rotators have been investigated in
\cite{Mao92,Sad95,Sad96,Mai97a} and turn out to be much more subtle 
than those of the basic vibrators.
The scenario is always a sequence of a tangent and a pitchfork bifurcation 
which occur at two nearby bifurcation energies.
As an example we discuss the creation of the closed orbit $R_3^1$ in 
a bifurcation out of the third repetition of the basic rotator $R_3$.
The closed orbit parameters are presented in figure~\ref{fig3}, which
shows the energy dependence of the action difference $\Delta\tilde S/2\pi$ 
with the scaled action of the orbit $R_3$ taken as the reference action, 
and the element $\tilde m_{12}$ of the scaled monodromy matrix.
Real and complex ``ghost'' orbits are marked by solid and dashed lines,
respectively.
The real closed orbits $R_3$ and $R_3^1$ collide in a pitchfork bifurcation
at scaled energy $\tilde E=-0.48284$.
Below that bifurcation energy a ghost orbit in the complex phase space,
albeit with real action and monodromy matrix, does exist.
This orbit participates in a tangent bifurcation at scaled energy
$\tilde E=-0.48477$.
Unlike a conventional tangent bifurcation, where two real orbits are
created out of ghost orbit predecessors \cite{Kus93}, the tangent
bifurcation shown in figure~\ref{fig3} possesses the peculiar property
that all participating orbits are complex ghosts.
One ghost orbit with complex action and monodromy matrix and its complex 
conjugate companion bifurcate at $\tilde E=-0.48477$ into two genuinely
different ghost orbits with real actions and monodromy matrices.
The significance of ghost orbit bifurcations in semiclassical spectra
has already been demonstrated in \cite{Bar99a,Bar99b}.
The closed orbit parameters, including those of the complex ghosts, are 
required for the construction of uniform semiclassical approximations 
in Sec.~\ref{uni:sec}.

A systematic survey of the bifurcations of the basic rotators reveals
that the bifurcation scenario with the ghost orbit tangent bifurcation 
as shown in figure~\ref{fig3} is restricted to bifurcations of the basic 
rotators at scaled energies $\tilde E<-0.418$ \cite{Fab02}.
At higher energies conventional tangent bifurcations are found instead of
ghost orbit bifurcations.
Examples are the creation of orbits $R_2^1$ at $\tilde E\approx -0.317$ 
\cite{Mai97a} or $R_3^2$ at $\tilde E\approx -0.209$.

\subsection{Bifurcations of orbits $V_\mu^\nu$ and $R_\mu^\nu$}
\label{V_R_bif:sec}
\begin{figure}
\begin{center}
\includegraphics[width=0.75\columnwidth]{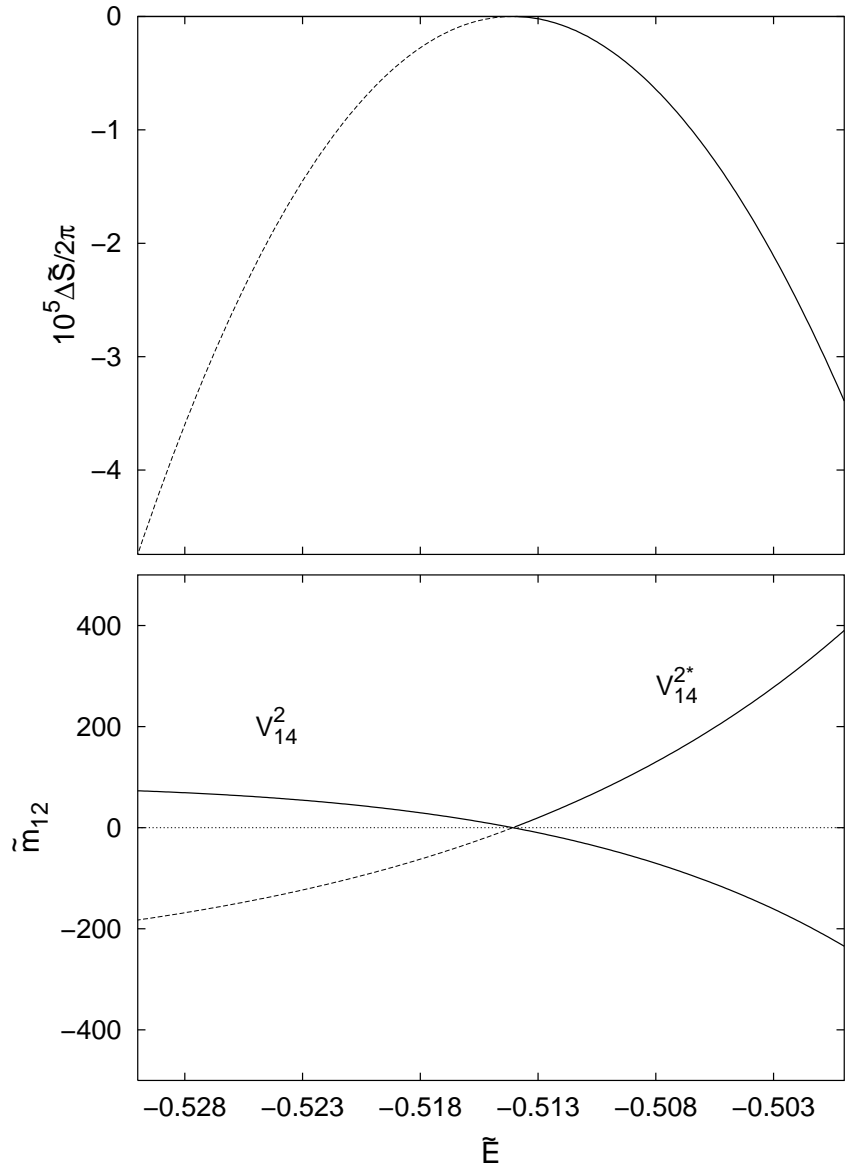}
\end{center}
\caption{Closed orbit parameters for the pitchfork bifurcation of
the vibrator $V_{14}^2$ at $\tilde E = -0.514$.
The real asymmetric orbit is labelled $V_{14}^{2\ast}$.
The predecessor ghost orbit is marked by dashed lines.}
\label{fig4}
\end{figure}

After being created, the vibrators $V_\mu^\nu$ and rotators $R_\mu^\nu$ 
can themselves undergo further bifurcations with increasing energy.
These are pitchfork bifurcations, where a pair of asymmetric orbits 
(with different initial and final angles, $\vartheta_i\ne\vartheta_f$) 
separate from a central symmetric obit with $\vartheta_i=\vartheta_f$
or $\vartheta_i=\pi-\vartheta_f$, which is real below and above the
bifurcation.
On the two asymmetric orbits the electron follows the same trajectory in 
different directions, i.e., the final angle of one orbit is the initial 
angle of the other orbit and vice versa.
Below the bifurcation point the asymmetric orbits become a pair of complex 
conjugate ghost orbits.
Again the electron follows the same trajectory in different directions 
and therefore must have real actions and monodromy matrices.
For illustration the closed orbit parameters of the first pitchfork
bifurcation of the orbit $V_{14}^2$ at $\tilde E=-0.514$ are
presented in figure~\ref{fig4}.
A typical feature of the pitchfork bifurcations is the nearly linear 
dependence of the monodromy matrix element $\tilde m_{12}$ and the 
nearly quadratic behaviour of the action difference $\Delta\tilde S$ 
around the bifurcation energy.
For pitchfork bifurcations of vibrators $V_\mu^\nu$ the action of the 
asymmetric orbits exceeds the action of the symmetric orbit.
Pitchfork bifurcations of the rotators $R_\mu^\nu$ show the opposite 
behaviour, i.e., the action of the symmetric orbit $R_\mu^\nu$ exceeds 
the action of the newly created asymmetric orbits.

\bigskip
The bifurcations discussed so far are sufficient to describe the
complete bifurcation tree of all closed orbits at scaled energies
$\tilde E<-0.5$ and with recurrence times $\tilde T/2\pi<12$.
However, at higher energies more types of bifurcations exist,
e.g., new closed orbits can be created in tangent bifurcations
(without an accompanying pitchfork bifurcation as for the bifurcations
of the perpendicular orbit discussed in Sec.~\ref{R_bif:sec}).
The simplest of these orbits, called $X_1$ in \cite{Hol88,Mai91,Mai97a},
is created at the scaled energy $\tilde E=-0.11544216$.
As the calculations of the semiclassical photoabsorption spectra 
in Sec.~\ref{high_res_spectra:sec} are restricted to energies $\tilde E<-0.5$ 
we need not discuss the tangent bifurcations and the corresponding 
uniform semiclassical approximations of such orbits in more detail.

\section{Closed orbit theory and uniform approximations}
\label{uni:sec}
Closed orbit theory \cite{Du88,Bog89} provides a semiclassical approximation
to photoabsorption spectra of atoms in external fields, where the electron is
excited from a low lying initial state $|\Psi_i\rangle$ to a final Rydberg
state $|\Psi_n\rangle$.
It is convenient to introduce the quantum mechanical response function 
\begin{equation}
   g^{\rm qm}(E)
 = -\frac{1}{\pi}\langle\Psi_i|DG_E^+D|\Psi_i\rangle
 = -\frac{1}{\pi}
   \sum_n\frac{|\langle\Psi_i|D|\Psi_n\rangle|^2}{E-E_n+{\rm i}\epsilon}
\label{g_qm}
\end{equation}
with $D$ the dipole operator, $E_n$ the eigenenergy of the eigenstate
$|\Psi_n\rangle$, and $G_E^+$ the retarded Green's function. 
From the response function (\ref{g_qm}) relevant physical data such as
the oscillator strength 
\begin{equation}
 f(E) = 2(E-E_i)\, {\rm Im}\, g^{\rm qm}(E)
\label{f:eq}
\end{equation}
are readily obtained.
The semiclassical approximation to the exact quantum response function 
(\ref{g_qm}) is given by closed orbit theory as a sum of a smooth background 
term and an oscillatory part
\begin{equation}
 g^{\rm osc}(E) = \sum_{\rm co}{\cal A}_{\rm co}(E)
   {\rm e}^{{\rm i}(S_{\rm co}(E)-\frac{\pi}{2}\mu_{\rm co})} \; ,
\label{g_sc}
\end{equation}     
where the sum is to be taken over all closed orbits (co) starting at and 
returning back to the nucleus.
$S_{\rm co}$ and $\mu_{\rm co}$ are the classical action and Maslov index
of the closed orbit, respectively.
The amplitudes ${\cal A}_{\rm co}$ depend on the symmetry of the orbits 
and read
\begin{equation}
   {\cal A}_{\rm co}^{\rm non}
 = 2(2\pi)^{3/2}\sqrt{\frac{\sin \vartheta_i \sin \vartheta_f}
   {|m_{12}|}} {\cal Y}(\vartheta_i){\cal Y}(\vartheta_f)
   {\rm e}^{{\rm i}\frac{\pi}{4}}
\label{amp_non}
\end{equation}
for non-axial closed orbits, which in the three-dimensional coordinate space
form a rotationally invariant family of orbits around the field axis, and
\begin{equation}
   {\cal A}_{\rm co}^{\rm ax}
 = \frac{4\pi}{|m_{12}|}{\cal Y}(\vartheta_i){\cal Y}(\vartheta_f)
\label{amp_ax}
\end{equation}
for axial orbits \cite{Gao92,Mai94}, i.e., the basic vibrators parallel 
to the magnetic field.
In Eqs.~(\ref{amp_non}) and (\ref{amp_ax}) $\vartheta_i$ and $\vartheta_f$
are the initial and final angle, and $m_{12}$ is an element of the monodromy
matrix of the closed orbit.
Note that these parameters depend on the energy $E$.
The functions ${\cal Y}(\vartheta)$ characterize the initial state 
$|\Psi_i\rangle$ and the polarization of the dipole transition, and are 
linear combinations of spherical harmonics $Y_{lm}(\vartheta,0)$.
The closed orbit amplitudes (\ref{amp_non}) and (\ref{amp_ax}) 
are valid in integrable as well as chaotic regimes.
By contrast, the trace formulas of periodic orbit theory are different for 
integrable \cite{Ber76} and chaotic \cite{Gut90} systems.
However, the closed orbits must be isolated, i.e., Eqs.~(\ref{amp_non}) and 
(\ref{amp_ax}) fail near bifurcations where different orbits approach 
each other and eventually collide.
The element $m_{12}$ of the monodromy matrix vanishes at bifurcations, and
the semiclassical amplitudes ${\cal A}_{\rm co}$ of the isolated closed
orbit contributions suffer from unphysical singularities.

To obtain a smooth contribution to the semiclassical response function
$g^{\rm osc}(E)$ where closed orbits bifurcate, uniform approximations are
needed. 
Their construction requires a detailed description of the bifurcation scenario.
In the language of catastrophe theory \cite{Pos78,Ber80}, this can be 
achieved in terms of normal forms whose stationary points correspond to the 
classical closed orbits \cite{Mai97a,Bar03b}.
The codimension of the bifurcation scenario coincides with the codimension 
of its normal form.
For a generic bifurcation, it is at most the number of external parameters, 
which is one in the case of the diamagnetic Kepler problem.
In this system two types of generic bifurcations exist: tangent bifurcations 
and pitchfork bifurcations.
They can be described by the fold and the symmetric cusp, respectively.
More complicated bifurcation scenarios are composed of several generic 
bifurcations and modelled by catastrophes of higher codimension.
If the individual bifurcations are closely spaced, it is important to 
construct a uniform approximation that describes them collectively.
Various uniform semiclassical approximations have already been constructed 
for the hydrogen atom in an electric \cite{Gao97,Sha98,Bar02,Bar03a} 
and a magnetic \cite{Mai97a,Mai98a} field.

As a starting point for catastrophes of corank one we consider the ansatz
\begin{equation}
   g^{\rm osc}_{\rm uni}(E)
 = \int p(t) {\rm e}^{{\rm i}\Phi_{\bf a}(t)}\,{\rm d} t \,
   {\rm e}^{{\rm i}(S_0 - \frac{\pi}{2}\nu_0)} \; ,
\label{uniansatz}
\end{equation}             
where $\Phi_{\bf a}(t)$ is the normal form of the catastrophe
depending on the parameters ${\bf a}=(a_1, a_2, \dots, a_k)$ with
$k$ being the codimension.
The uniform approximation (\ref{uniansatz}) is supposed to reproduce the
closed orbit sum of all orbits participating in the bifurcation scenario if
the distance from the bifurcation is large.
$S_0$ is an energy dependent reference action, e.g., the action of a 
central closed orbit.
The integer $\nu_0$ and the function $p(t)$ must be chosen to asymptotically 
provide the correct phase and amplitude of the uniform approximation.
The classical actions of the closed orbits contributing to the bifurcation
scenario are related to the stationary values of the normal form 
$\Phi_{\bf a}(t)$.
To determine the parameters of the normal form $\Phi_{\bf a}(t)$ and the 
amplitude function $p(t)$ we use the asymptotic expressions of the
uniform approximation (\ref{uniansatz}) far away from the bifurcations,
where the integral can be evaluated in stationary phase approximation.
The stationary phase (sp) method applied to equation~(\ref{uniansatz}) yields
\begin{equation}
 g_{\rm uni}^{\rm osc}(E)\stackrel{\mbox{\scriptsize sp}}{\approx}
 \sum_n\frac{\sqrt{2\pi {\rm i}}\; p(t_n)} {\sqrt{|\Phi''(t_n)|}} \,
 {\rm e}^{-{\rm i}\frac{\pi}{2}(\nu_0+\nu_n)}
 {\rm e}^{{\rm i}(S_0+\Phi(t_n))} \; ,
\label{g_uni}
\end{equation}
where the $t_n$ are the stationary points of $\Phi_{\bf a}(t)$.
The constant integer $\nu_0$ is given by the Maslov indices 
of the closed orbits, which change at the bifurcations, i.e.,
$\mu_n=\nu_0+\nu_n$, with
\begin{equation}
 \nu_n = \left\{ \begin{array}{ll}1, & \Phi_{\bf a}''(t_n)<0,\\ 
                                  0, & \mbox{otherwise.} \end{array} \right.
\end{equation}
The sum over the stationary points in equation~(\ref{g_uni}) is identified
with the sum (\ref{g_sc}) over the closed orbits colliding in the bifurcation
scenario described by the normal form $\Phi_{\bf a}(t)$.
From the comparison of the two equations (\ref{g_sc}) and (\ref{g_uni}) 
we obtain the conditions
\begin{equation}
 S_0 + \Phi(t_n) = S_n
\label{gl.wirk0}
\end{equation}
and
\begin{equation}
 \frac{\sqrt{2\pi {\rm i}}\;p(t_n)}{\sqrt{|\Phi''(t_n)|}} = \mathcal{A}_n \; .
\label{gl.ampl}
\end{equation} 
Eqs.~(\ref{gl.wirk0}) and (\ref{gl.ampl}) are valid not only for 
real orbits but hold also for complex ghost orbits with the slight 
modifications \cite{Scho97} that $|m_{12}|$ and $|\Phi''(t_n)|$ are 
replaced with $\mbox{sign}(\mbox{Re}\, m_{12})\,m_{12}$ and
$\mbox{sign}(\mbox{Re}\,\Phi''(t_n))\,\Phi''(t_n)$, respectively.
These equations are now used to determine, from the numerically calculated
closed orbit parameters $S_n$ and ${\cal A}_n$, the parameters of the normal
form $\Phi_{\bf a}(t)$ and the amplitude functions $p(t)$ for various types
of catastrophes.
The functions $\Phi_{\bf a}(t)$ and $p(t)$ are then inserted in 
equation~(\ref{uniansatz}) to obtain the uniform approximations.
In the following we discuss the butterfly catastrophe related to bifurcations
of the perpendicular orbit, the symmetric cusp catastrophe related to
bifurcations of the rotators and vibrators, and the uniform approximations 
for bifurcations of the parallel orbit.

\subsection{Uniform approximations for bifurcations of the perpendicular orbit}
The bifurcation scenarios of the perpendicular orbit discussed 
in Sec.~\ref{R_bif:sec} are described by the codimension-2 symmetric 
butterfly catastrophe \cite{Mai97a}.
A local approximation, which removes the singularities at the bifurcation
energies but is not valid at energies far away from the bifurcation points 
has been presented in \cite{Mai97a}.
Here a uniform approximation is derived that removes the unphysical
divergences at the bifurcations and agrees asymptotically with the 
closed orbit sum of the isolated orbits.

\begin{figure}
\begin{center}
\includegraphics[width=0.75\columnwidth]{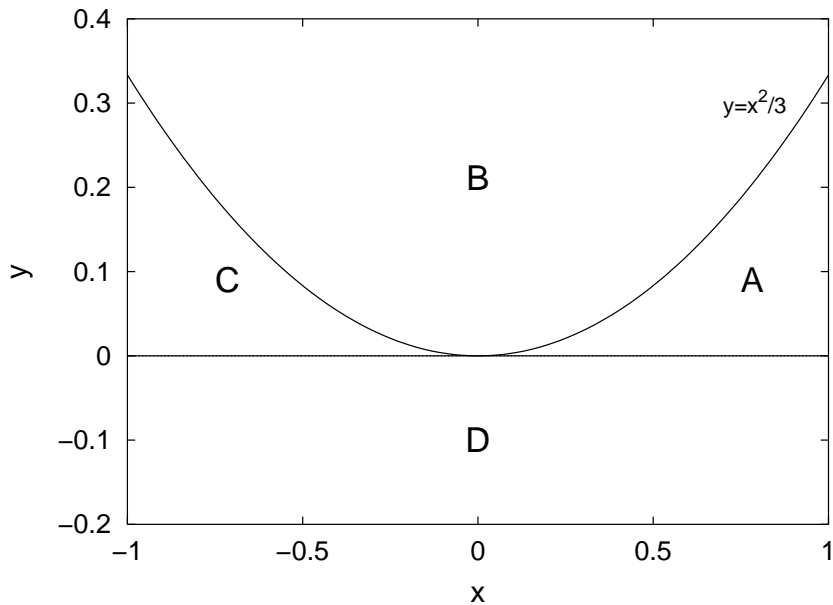}
\end{center}
\caption{Characterization of the stationary points of the normal form
$\Phi(t)$ in equation~(\ref{phi}).  The functions $y=0$ and $y=x^2/3$ divide
the $(x,y)$-plane into different domains.
A: $t_{1\dots 4}\in {\rm i}\mathbb{R}$;
B: $t_{1\dots 4}\in\mathbb{C}$;
C: $t_{1\dots 4}\in\mathbb{R}$;
D: $t_{1,2}\in\mathbb{R}$, $t_{3,4}\in {\rm i}\mathbb{R}$.}
\label{fig5}
\end{figure}

The normal form of the symmetric butterfly catastrophe reads
\begin{equation}
 \Phi(t)= -\left(t^6+xt^4+yt^2\right) \; ,
\label{phi}
\end{equation}
where the two real unfolding parameters $x$ and $y$ depend on the energy and
magnetic field strength and must be determined from equation~(\ref{gl.wirk0}).
To that end, the stationary points of $\Phi(t)$ at
\begin{equation}
 t_0=0 \; , \quad
 t_{1,2} = \pm\sqrt{(-x+\delta)/3} \; , \quad
 t_{3,4} = \pm\sqrt{(-x-\delta)/3} \; ,
\label{t_n}
\end{equation}
with $\delta\equiv\sqrt{x^2-3y}$ are identified with the (real or complex)
closed orbits contributing to the bifurcation scenario:
Real stationary points are real orbits, purely imaginary stationary points 
are ghost orbits with real action, and complex points are ghost orbits with 
complex action.
The trivial stationary point $t_0$ is the perpendicular orbit that is
real on both sides of the bifurcation.
The nontrivial stationary points are real, imaginary, or general complex
numbers in various regions of the parameters $x$ and $y$ as illustrated 
in figure~\ref{fig5}.
The lines $y=0$ and $y=x^2/3$ divide the $(x,y)$-plane into different
domains and characterize the parameter values where pitchfork or tangent
bifurcations, respectively, of the closed orbits occur. The stationary
points $t_{1\dots 4}$ are imaginary in region A, complex in region B and
real in region C. In region D, $t_{1,2}$ are real whereas $t_{3,4}$ are
imaginary.

\begin{figure}
\begin{center}
\includegraphics[width=\columnwidth]{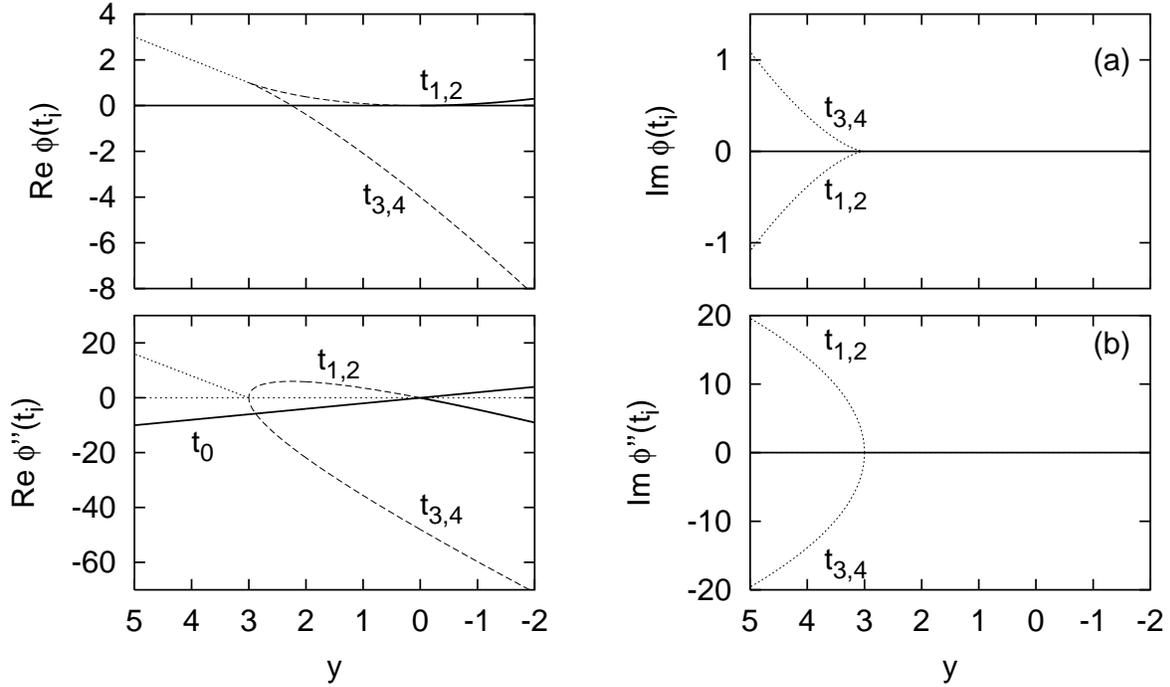}
\end{center}
\caption{(a) Stationary values of the normal form $\Phi(t)$ in 
equation~(\ref{phi}) as function of the unfolding parameter $y$ ($x=3>0$).
(b) Second derivative $\Phi''(t)$ at the stationary points.
The qualitative agreement with figure~\ref{fig3} is evident.}
\label{fig6}
\end{figure}

These observations translate into the following bifurcation scenarios: 
In region \textrm{D} the stationary points $t_{1,2}$ represent the real 
orbits $R_\mu^\nu$.
If $x>0$, as $y$ is increased these orbits collide with the perpendicular 
orbit ($t_0$) at $y=0$ and become imaginary, corresponding to ghost orbits 
with real action, in region \textrm{A}.
At $y=x^2/3$ they collide with $t_{3,4}$, and finally in region \textrm{B} 
the stationary points $t_{1\dots 4}$ represent a quadruple of ghost orbits 
with complex action.
This scenario with unfolding parameter $x>0$ is observed for bifurcations
at scaled energies $\tilde E<-0.418$.
The scenario for $x<0$ is similar, the difference being that all stationary 
points and corresponding closed orbits are real in region \textrm{C}.
This scenario has been observed at scaled energies 
$\tilde E>-0.418$ \cite{Fab02}.
The stationary values of the normal form $\Phi(t)$ in equation~(\ref{phi})
and the second derivative $\Phi''(t)$ as functions of the unfolding 
parameter $y$ are presented in figure~\ref{fig6}.
Evidently, figure~\ref{fig6} qualitatively agrees with the closed orbit 
parameters $\Delta\tilde S$ and $\tilde m_{12}$ shown in figure~\ref{fig3}.

The unfolding parameters $x$ and $y$ can now be determined from the actions
of the colliding orbits using equation~(\ref{gl.wirk0}).
When $S_0$ is identified with the action of the perpendicular orbit, we obtain
\begin{eqnarray}
\label{gl.wirk1}
    \Phi(t_1) &\equiv& \Phi_1
 = -\frac{2}{27}(x-\delta)\delta^2+\frac{1}{9}xy = S_1-S_0 = \Delta S_1 \; ,\\
\label{gl.wirk3}
    \Phi(t_3) &\equiv& \Phi_3
 = -\frac{2}{27}(x+\delta)\delta^2+\frac{1}{9}xy = S_3-S_0 = \Delta S_3 \; ,
\end{eqnarray}
where $S_1$ and $S_3$ denote the actions of the corresponding orbits
described above.
The sum and the difference of Eqs.~(\ref{gl.wirk1}) and (\ref{gl.wirk3}) 
yields
\begin{eqnarray}
\label{gl.delta1}
  \Delta S_1-\Delta S_3 &=& \Phi_1-\Phi_3 = \frac{4}{27}\delta^3 \; ,\\
\label{gl.x}
  \Delta S_1+\Delta S_3 &=& \Phi_1+\Phi_3 = \frac{2}{27}
      \left(x^3-3\delta^2x\right) \; .
\end{eqnarray}
Now 
\begin{equation}
 \delta\equiv\sqrt{x^2-3y} = 3\sqrt[3]{(\Phi_1-\Phi_3)/4}
\end{equation}
follows from equation~(\ref{gl.delta1}), and 
the unfolding parameter $x$ is obtained from the solution of the 
cubic equation (\ref{gl.x}) using Cardano's formula as \cite{Abr65}
\begin{equation}
 x = \lambda\,\sqrt[3]{\frac{27}{4}\left[(\Phi_1+\Phi_3)
      +2\sqrt{\Phi_1\Phi_3}\right]}
   + \lambda^\ast\,\sqrt[3]{\frac{27}{4}\left[(\Phi_1+\Phi_3)
      -2\sqrt{\Phi_1\Phi_3}\right]} \; ,
\label{xlsg}
\end{equation}
with $\lambda\in\{1,(-1\pm {\rm i}\sqrt{3})/2\}$.
The parameter $x$ becomes a smooth function of the energy with the choice
\begin{equation}
 \lambda = \left\{\begin{array}{cl}1, & \tilde E < \tilde E_c\; , \\
     -(1+{\rm i}\sqrt{3})/2, & \tilde E > \tilde E_c\; , \end{array}\right.
\label{lambda}
\end{equation}
where $\tilde E_c$ is the bifurcation energy of the pitchfork bifurcation,
i.e.\ $\Phi_1=0$.
Finally the unfolding parameter $y$ is given as $y=(x^2-\delta^2)/3$.

The next step is to construct the function $p(t)$ in the uniform
approximation (\ref{uniansatz}) in such a way that equation~(\ref{gl.ampl}) 
is valid at the stationary points $t_n$ of the normal form $\Phi(t)$.
There is considerable freedom in doing so, and one will strive for an ansatz 
for $p(t)$ that is as simple as possible.
In all cases discussed in the literature so far 
(e.g., \cite{Scho97,Sie98,Sha98,Bar03b}), it has been found sufficiently 
accurate to choose $p(t)$ to be a low-order polynomial with as many 
undetermined coefficients as there are conditions imposed by~(\ref{gl.ampl}).
For the bifurcation scenario described by~(\ref{phi}), however, we find that 
a polynomial ansatz for $p(t)$ yields a uniform approximation that in the 
bifurcation region differs from the expected results and assumes the correct
asymptotic behaviour only at huge distances from the bifurcation. 
This scenario thus calls for a more thorough analysis of the amplitude 
function.

Because contributions to the integral in the uniform 
approximation~(\ref{uniansatz}) arise chiefly in a neighbourhood of the
origin, where the stationary points of the normal form $\Phi(t)$ are
located, the amplitude function $p(t)$ need only be accurate in that
region.
A polynomial ansatz is justified if $p(t)$ is nearly constant in
the region of interest, so that the $t$-dependent terms are small
correction to the constant term and high-order terms that are not included
in the ansatz are negligible.
A comparison of figure~\ref{fig3} and figure~\ref{fig6} reveals that the 
energy-dependence of the monodromy matrix element $m_{12}$ is well described 
by the second derivative of the normal form.
To satisfy~(\ref{gl.ampl}), it remains for $p(t)$ to describe the energy 
dependence of the amplitude through the angles $\vartheta_i$ and $\vartheta_f$.
These considerations suggest the ansatz
\begin{equation}
 p(t) = (at^4+bt^2+c)\, {\cal Y}^2(\vartheta (t)) \; ,
\label{gl.p_t}
\end{equation}
with the same angular function ${\cal Y}(\vartheta)$ as in 
equation~(\ref{amp_non}).
The mapping $\vartheta(t)$ from the normal form coordinate $t$ to the 
angle $\vartheta$ is in turn modelled by the polynomial ansatz
\begin{equation}
 \vartheta(t) = \frac{\pi}{2}+vt+ut^3 \; ,
\end{equation}
with the coefficients $u$ and $v$ chosen such that $\vartheta(t)$ maps the
stationary points of the normal form~(\ref{phi}) onto the numerically
determined angles $\vartheta_i$ and $\vartheta_f$.
(Note that $\vartheta_i=\vartheta_f$ or $\vartheta_i=\pi-\vartheta_f$ for 
all orbits involved.)

Inserting equation~(\ref{gl.p_t}) into (\ref{gl.ampl}), we obtain the three
coefficients $a,b,c$ in (\ref{gl.p_t}) as
\begin{eqnarray}
\label{gl.a}
a &=&
 \left( \frac{3\mathcal{A}^*_0}{y}\sqrt{\frac{|y|}{2\pi}}
 -\frac{3x}{2y\delta} \sqrt{\frac{2}{3\pi}}
 \left[\mathcal{A}^*_1\sqrt{\eta_1\delta(x-\delta)}
  -\mathcal{A}^*_3\sqrt{\eta_3 \delta(x+\delta)}\right]\right. \nonumber\\
 &-& \left. \frac{3}{2y}
 \sqrt{\frac{2}{3\pi}}\left[\mathcal{A}^*_1\sqrt{\eta_1 \delta(x-\delta)}
 +\mathcal{A}^*_3\sqrt{\eta_1 \delta(x+\delta)}\right]\right)(1-{\rm i}) \; ,\\
\label{gl.b}
b &=&
 \left(\frac{x\mathcal{A}^*_0}{y}\sqrt{2\frac{|y|}{\pi}}
 -\frac{x}{y}\sqrt{\frac{2}{3\pi}}\left[\mathcal{A}^*_1
 \sqrt{\eta_1 \delta(x-\delta)}  
 +\mathcal{A}^*_3\sqrt{\eta_3 \delta(x+\delta)}\right]\right. \nonumber\\
 &-&\left.\left(\frac{3}{2\delta}+\frac{\delta}{y}\right)\sqrt{\frac{2}{3\pi}}
 \left[\mathcal{A}^*_1\sqrt{\eta_1 \delta(x-\delta)}
 -\mathcal{A}^*_3\sqrt{\eta_3 \delta(x+\delta)}\right]
 \right)(1-{\rm i}) \; , \\
\label{gl.c}
c &=& \mathcal{A}^*_0\sqrt{\frac{|y|}{2\pi}}(1-{\rm i}) \; ,
\end{eqnarray}
where ${\cal A}^\ast_n={\cal A}_n/({\cal Y}(\vartheta_i){\cal Y}(\vartheta_f))$
denotes the semiclassical amplitudes in equation~(\ref{amp_non}) without the 
angular functions, $x$, $y$ and $\delta$ are the parameters as introduced 
above, and $\eta_n=\mbox{sign}\,({\rm Re}\,\Phi''(t_n))$.
Note that all coefficients in the normal form $\Phi(t)$ and the amplitude 
function $p(t)$ are now given as explicit functions of the closed orbit 
parameters of the real and complex (ghost) orbits involved in the bifurcation
scenario.

\begin{figure}
\begin{center}
\includegraphics[width=0.8\columnwidth]{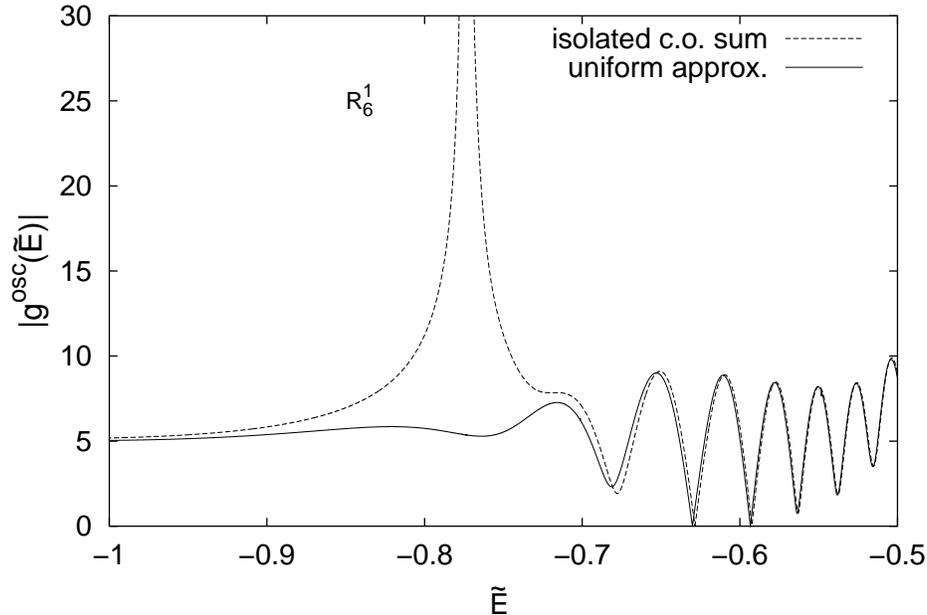}
\end{center}
\caption{Absolute value of the semiclassical response function for the 
bifurcation scenario of orbits $R_6$ and $R_6^1$.
Dashed line: isolated closed orbit sum;
solid line: uniform approximation.}
\label{fig7}
\end{figure}

With the normal form $\Phi(t)$ and the amplitude function $p(t)$ at hand,
it is possible to evaluate the uniform approximation (\ref{uniansatz}).
The integral must be solved numerically.
For $t\to\pm\infty$ the integrand is highly oscillating and must therefore
be regularized by multiplication with a factor of the form 
$\exp(-\epsilon t^m)$ with the small $\epsilon>0$ and the power $m>0$
chosen appropriately.
As an example figure~\ref{fig7} presents the absolute value of the
semiclassical response function for the orbits $R_6$, $R_6^1$, and
the ghost orbits associated in the bifurcation scenario.
The isolated closed orbit sum (dashed line) suffers from the unphysical
divergence around $\tilde E=-0.78$.
By contrast, the uniform approximation (solid line) is a smooth function 
at all energies.
The modulations of the amplitude at $\tilde E\gtrsim -0.7$ are caused by
the interference of the real closed orbits $R_6$ and $R_6^1$.
Note that the uniform approximation at large distances from the bifurcation 
energies asymptotically agrees with the isolated closed orbit sum.

\subsection{Uniform approximations for bifurcations of the rotators 
and vibrators}
The pitchfork bifurcations of the rotators $R_\mu^\nu$ and vibrators
$V_\mu^\nu$ discussed in Sec.~\ref{V_R_bif:sec} are described by the 
normal form of the symmetrized cusp catastrophe 
\begin{equation}
 \Phi_a(t) = \frac{1}{4}t^4-\frac{1}{2} a t^2 \; ,
\label{Phicusp}
\end{equation} 
which has one unfolding parameter $a$.
The stationary points at $t=0$  and $t=\pm \sqrt{a}$ correspond to the
symmetric (sym) and asymmetric (asym) orbits, respectively. 
The normal form parameter 
\begin{equation}
 a = \pm 2 \sqrt{S_{\rm sym}-S_{\rm asym}}
\end{equation} 
is a function of the actions of the closed orbits involved.
It has to be chosen positive if the asymmetric orbits are real, and 
negative otherwise.

Due to symmetry properties of the orbits the amplitude function $p(t)$ in 
equation~(\ref{uniansatz}) must be an even function of $t$.
It is sufficient to use a simple polynomial ansatz $p(t) = p_0 + p_2 t^2$.
The coefficients are obtained as \cite{Bar03b}
\begin{eqnarray}
\label{p0:eq}
 p_0 &=& \sqrt{\frac{|a|}{2\pi}}{\cal A}_{\rm sym}
     {\rm e}^{-{\rm i}\frac{\pi}{4}} \; , \\
\label{p2:eq}
 p_2 &=& \frac{1}{2a}\sqrt{\frac{|a|}{\pi}}\left({\cal A}_{\rm asym}
     -\sqrt{2}{\cal A}_{\rm sym}\right)
     {\rm e}^{-{\rm i}\frac{\pi}{4}} \; ,
\end{eqnarray}
with ${\cal A}_{\rm sym}$ and ${\cal A}_{\rm asym}$ the amplitudes 
of the (isolated) symmetric and asymmetric closed orbits as defined in 
equation~(\ref{amp_non}).
(Equations~(\ref{p0:eq}) and (\ref{p2:eq}) slightly differ from formulae
given in \cite{Bar03b} due to a different handling of the Maslov phase
in equations~(\ref{g_sc}) and (\ref{amp_non}).)
The uniform approximation can now be written as
\begin{equation}
   g^{\rm osc}_{\rm uni}
 = \int p(t) {\rm e}^{{\rm i} \Phi_a(t)} {\rm d} t\,
    {\rm e}^{{\rm i}(S_{\rm sym}-\frac{\pi}{2}\nu_0)}
 = (p_0I_0+p_2I_2) {\rm e}^{{\rm i} (S_{\rm sym}-\frac{\pi}{2}\nu_0)}
\label{uni_cusp}
\end{equation}
with the integrals 
$I_0\equiv\int \exp(i\Phi_a(t)){\rm d} t$ and
$I_2\equiv\int t^2 \exp(i\Phi_a(t)){\rm d} t$.
The integrals can be evaluated analytically in terms of Bessel functions 
$J_\nu(z)$ \cite{Abr65} and read \cite{Bar03b}
\begin{eqnarray}
\fl
 I_0 &=& \frac{\pi}{2}\sqrt{|a|}e^{-ia^2/8} \left[e^{i\pi/8}
   J_{-1/4}\left(\frac{a^2}{8}\right) +{\rm sign}\, a\, 
   e^{-i \pi/8} J_{1/4}\left(\frac{a^2}{8}\right)\right] \; , \\
\fl
 I_2 &=& i\pi\sqrt{|a|}e^{-ia^2/8}\left\{\left(\frac{1}{2a}-i\frac{a}{4}\right)
   \left[e^{i\pi/8}J_{-1/4}\left(\frac{a^2}{8}\right) +{\rm sign}\, a \,
   e^{-i\pi/8}J_{1/4}\left(\frac{a^2}{8}\right) \right] \right. \nonumber \\
\fl
   &+& \left. \frac{a}{8}e^{i \pi/8} \left[J_{-5/4}\left(\frac{a^2}{8}\right)
     + J_{3/4}\left(\frac{a^2}{8}\right) \right]
     + {\rm sign}\, a \, e^{-i\pi/8} \left[J_{-3/4}\left(\frac{a^2}{8}\right)
     - J_{5/4}\left(\frac{a^2}{8}\right) \right] \right\} .
\end{eqnarray}
The normal form (\ref{Phicusp}) and thus the uniform approximation
(\ref{uni_cusp}) describe the pitchfork bifurcation when the action of 
the symmetric orbit is larger than the action of the asymmetric orbits 
in the vicinity of the bifurcation.
This is true for the rotators $R_\mu^\nu$. 
For the vibrators $V_\mu^\nu$ the converse is true, i.e., close to the 
bifurcation the action of the asymmetric orbits exceeds the action of the 
symmetric orbit $V_\mu^\nu$.
In this case, called a dual cusp \cite{Pos78}, the normal form $\Phi_a(t)$ 
must be replaced with $-\Phi_a(t)$, which changes the sign of the 
stationary values.
The uniform approximation for the dual cusp is obtained by replacing the 
integrals $I_0$ and $I_2$ in equation~(\ref{uni_cusp}) with its complex 
conjugate, i.e., $g^{\rm osc}_{\rm uni}=(p_0I_0^*+p_2I_2^*)
\exp({\rm i}(S_{\rm sym}-\frac{\pi}{2}\nu_0))$.

\subsection{Uniform approximations for bifurcations of the parallel orbit}
In bifurcations of orbits parallel to the external field a rotationally
symmetric bundle of non-axial orbits splits from the axial orbit. 
This is true for both an external magnetic and an electric field.
The bifurcations of the axial orbits are described by the normal form 
$\Phi_a(t)=\frac{1}{4}t^4-\frac{1}{2}at^2$ which is formally the 
symmetric cusp, equation~(\ref{Phicusp}), but with the difference that $t$
is interpreted as a radial coordinate, $t=\sqrt{x^2+y^2}$.
For the amplitude function in (\ref{uniansatz}) the ansatz 
$p(t)=p_0+p_1(t^2-a)$ is used \cite{Bar03a}.
The stationary points at $t=\sqrt{a}$ describing the bundle of non-axial 
orbits lie on a circle with centre at $t=0$, which is the stationary point
describing the isolated axial orbit.
The parameter $a=\pm 2\sqrt{S_{\rm ax}-S_{\rm non}}$ is related to the 
classical action of the orbits, with positive and negative values 
of $a$ referring to real and ghost orbits, respectively.

The bifurcations of the ``uphill'' and ``downhill'' orbits parallel and
antiparallel to the electric field axis in the Stark system have already
been investigated and a uniform approximation for the creation or 
destruction of the non-axial orbits from the axial orbits has been 
constructed \cite{Gao97,Sha98}.
The uniform approximation is valid also to describe the bifurcations of
the basic vibrators in a magnetic field, and can be written in the concise 
form \cite{Bar02,Bar03a}
\begin{equation}
   g^{\rm osc}_{\rm uni}
 = \left[\frac{{\cal A}_{\rm non}}{1+{\rm i}}I(a)
    +\frac{{\rm i}}{a} \left(-|a|{\cal A}_{\rm ax}
    + \frac{1-{\rm i}}{\sqrt{2 \pi}} {\cal A}_{\rm non} \right) \right]
    e^{{\rm i} (S_{\rm ax}-\frac{\pi}{2}\nu_0)}
\label{g_uni_parallel}
\end{equation}
with
\begin{equation}
 I(a) = {\rm e}^{-{\rm i} a^2/4}\left[\frac{1+{\rm i}}{2}
      -C \left(\frac{-a}{\sqrt{2 \pi}}\right)
      - {\rm i} S\left(\frac{-a}{ \sqrt{2 \pi}}\right) \right]
\end{equation}
given in terms of the Fresnel integrals $C(z)$ and $S(z)$ \cite{Abr65}.
(Similar as above equation~(\ref{g_uni_parallel}) slightly differs from 
the result given in \cite{Bar02,Bar03a} due to a different handling of 
the Maslov phase in equations~(\ref{g_sc})--(\ref{amp_ax}).)

\section{High resolution photoabsorption spectra}
\label{high_res_spectra:sec}
With the closed orbit theory and the uniform approximations at hand 
we can now obtain the semiclassical response function $g^{\rm sc}(E)$ 
via summation of the closed orbit contributions.
In the vicinity of bifurcations the contributions of isolated orbits must
be replaced with the uniform approximations.
The semiclassical photoabsorption spectrum is then readily given by
equation~(\ref{f:eq}) with $g^{\rm qm}(E)$ replaced with its semiclassical
analogue.
However, the closed orbit sum diverges when it is extended over all closed
orbits.
When it is truncated, e.g., by neglecting orbits with recurrence time 
$T>T_{\max}$, it yields only low resolution spectra.
To obtain high resolution spectra, i.e., discrete eigenenergies $E_n$ and 
individual transition matrix elements 
$d_n=| \langle \Psi_i|D|\Psi_n \rangle |^2$,
we adopt the harmonic inversion method \cite{Mai99d} which has been
successfully applied in semiclassical mechanics, either to extract the
actions and amplitudes of classical orbits from quantum spectra \cite{Mai97b}
or to calculate quantum mechanical quantities from classical orbits 
\cite{Mai97c,Mai98b,Mai99a,Mai99b}.
It has also been demonstrated that harmonic inversion is a powerful tool 
for semiclassical quantization using bifurcating orbits \cite{Bar02,Bar03a}.
To keep our presentation self-contained, we briefly outline the basic ideas.

\subsection{The harmonic inversion method}
\label{hi:sec}
In a first step, both the quantum (\ref{g_qm}) and the semiclassical 
response function (\ref{g_sc}) -- the smooth part can be neglected -- are
Fourier transformed into time domain.
The Fourier integrals are restricted to the energy window 
$[E_{\min},E_{\max}]$ where the closed orbit parameters have been 
calculated.
The windowed Fourier transforms result in the band-limited time signals
\begin{eqnarray}
\label{C_qm}
 C^{\rm qm}(t) &=& -\frac{1}{2\pi^2}\int_{E_{\min}}^{E_{\max}}
   \sum_n\frac{d_n}{E-E_n+{\rm i}\epsilon} {\rm e}^{-{\rm i} Et} {\rm d} E =
  \frac{{\rm i}}{\pi}\sum_n d_n {\rm e}^{-{\rm i} E_nt} \; , \\
\label{C_sc}
 C^{\rm sc}(t) &=& \frac{1}{2 \pi}\int_{E_{\min}}^{E_{\max}}
 \sum_{\rm co}{\mathcal A}_{\rm co}(E)
 {\rm e}^{{\rm i} S_{\rm co}(E)}{\rm e}^{-{\rm i} Et}{\rm d} E \; .
\end{eqnarray}
In the quantum signal (\ref{C_qm}) the sum is restricted to the
eigenenergies $E_n$ in the range $E_{\min}<E_n<E_{\max}$, i.e.,
only a relatively small number of parameters $\{E_n,d_n\}$ must be 
determined if the energy window is chosen appropriately.
In the semiclassical signal (\ref{C_sc}) only those closed orbits
contribute within a stationary phase approximation whose recurrence 
times $T$ are less than the total length $T_{\max}$ of the time signal.
This means that the semiclassical signal (\ref{C_sc}) can be constructed
if the set of the closed orbits with recurrence times $T<T_{\max}$ is
known in the energy interval $E_{\min}<E<E_{\max}$.
The semiclassical eigenenergies and transition matrix elements are now
obtained, in the second step, by adjusting the semiclassical signal 
(\ref{C_sc}) to its quantum analogue (\ref{C_qm}) with the $\{E_n,d_n\}$
being free adjustable parameters.
The technical details to solve this nonlinear fit problem are given 
in Ref.~\cite{Mai00}.

The required signal length $T_{\max}$ to achieve convergence of the 
harmonic inversion procedure depends on the mean level spacing 
$\bar\varrho(E)$ in the energy range $[E_{\min},E_{\max}]$ 
and reads $T_{\max}>4\pi\bar\varrho(E)$ \cite{Mai99d}.
The efficiency of the quantization method can be improved by using a
cross-correlated semiclassical recurrence signal \cite{Mai99a,Mai99c,Mai99d}.
The idea is to use a set of $L$ independent initial states $|\Psi_i\rangle$
and to construct the cross-correlated response function
\begin{equation}
 g^{\rm qm}_{ij}(E) = -\frac{1}{\pi} \langle \Psi_i|D G_E^+ D|\Psi_j \rangle
   \quad ; \quad i,j=1,2,\dots,L \, .
\end{equation}
Application of the windowed Fourier transform as in Eqs.~(\ref{C_qm}) and 
(\ref{C_sc}) yields the quantum $L\times L$ cross-correlated time signal
\begin{equation}
   C^{\rm qm}_{ij}
 = \frac{{\rm i}}{\pi} \sum_n b_{in} b_{jn} {\rm e}^{-{\rm i} E_nt} \; ,
\label{CqmX}
\end{equation}
with $b_{in}=\langle\Psi_i|D|\Psi_n\rangle$, and its semiclassical analogue
\begin{equation}
 C^{\rm sc}_{ij}(t) = \frac{1}{2 \pi}\int_{E_{\min}}^{E_{\max}}
  \sum_{\rm co}{\mathcal A}_{{\rm co},ij}(E){\rm e}^{{\rm i} S_{\rm co}(E)}
  {\rm e}^{-{\rm i} Et} {\rm d} E \; .
\label{CscX}
\end{equation}
For the various initial states $|\Psi_i\rangle$ the amplitudes 
${\mathcal A}_{{\rm co},ij}(E)$ in equation~(\ref{CscX}) differ by the use 
of various angular functions ${\cal Y}(\vartheta)$ in the closed orbit 
amplitudes (\ref{amp_non}) \cite{Mai99a}.
The semiclassical cross-correlated time signal (\ref{CscX}) can be
adjusted to its quantum form (\ref{CqmX}) with the $\{E_n,b_{in}\}$ being
the adjustable parameters by an extension of the harmonic inversion
method to cross-correlated time signals \cite{Wal95,Man98,Mai99d}.
The idea is to identify the cross-correlated recurrence function 
$C^{\rm sc}_{ij}(t)$, which is known on an equidistant time grid $t=n\tau$, 
with the cross-correlated time signal
$C_{ij}(n\tau)=(\Phi_i|\exp(-i n\tau\hat H_{\rm eff})| \Phi_j)$ of an 
effective Hamiltonian $\hat H_{\rm eff}$ with the (not explicitly known) 
states $|\Phi_i)$ and $|\Phi_j)$.
The operator $\hat H_{\rm eff}$ with eigenvalues $E_n$ is complex symmetric,
and $(x|y)$ denotes a complex symmetric (not Hermitian) inner product.
In an appropriate basis set the problem of extracting the $\{E_n,b_{in}\}$
can be reformulated as a generalized eigenvalue problem where all matrix
elements can be expressed in terms of the time signal $C_{ij}(n\tau)$.

The advantage of using the cross-correlation approach can be understood 
based on the argument that the total amount of independent information 
contained in the $L\times L$ signal is $L(L+1)$ multiplied by
the length of the signal, while the total number of unknowns
$\{E_n,b_{in}\}$ is $(L+1)$ times the total number of poles $E_n$.
Therefore the informational content of the $L\times L$ signal per unknown
parameter is increased, compared to the case of equation~(\ref{C_sc}), by a
factor of $L$.  This means that the required signal length
$T_{\max}=4\pi\bar\varrho(E)$ for a one-dimensional recurrence signal is
reduced by about a factor of $L$ for an $L\times L$ cross-correlated
recurrence signal.

In Sec.~\ref{results:sec} we investigate dipole transitions from the 
initial state $|\Psi_1\rangle = |2p0\rangle$ with light polarized parallel 
to the magnetic field axis to final states with magnetic quantum number $m=0$.
For this transition the angular function in equation~(\ref{amp_non}) 
reads \cite{Mai99d}
\begin{equation}
 {\cal Y}_1(\vartheta) = \frac{1}{\sqrt{2\pi}}2^7e^{-4}
   \left(4\cos^2\vartheta -1\right) \; .
\end{equation} 
Results are obtained by harmonic inversion of a one-dimensional and a
$2\times 2$ cross-correlated recurrence signal.
For the construction of the $2 \times 2$ cross-correlated signal we use 
for simplicity as a second transition formally an outgoing $s$-wave, viz.\
$D|\Psi_2\rangle \propto Y_{00}$, and, thus, ${\cal Y}_2(\vartheta)=1$.

\subsection{Results and discussion}
\label{results:sec}
For the semiclassical photoabsorption spectra we calculated all closed
orbits with recurrence times $\tilde T < \tilde T_{\max}=73.5$
in the scaled energy range $\tilde E\in [-1,-0.5]$ on an 
equidistant energy grid with step width $\Delta\tilde E=10^{-4}$.
The closed orbits have been used to construct low-resolution photoabsorption
spectra by superimposing the semiclassical contributions of the isolated 
orbits or, close to bifurcations, the uniform approximations.
High-resolution photoabsorption spectra with individual semiclassical
eigenenergies and transition matrix elements are obtained by the harmonic
inversion method as explained in Sec.~\ref{hi:sec}.

\begin{figure}
\begin{center}
\includegraphics[width=.9\columnwidth]{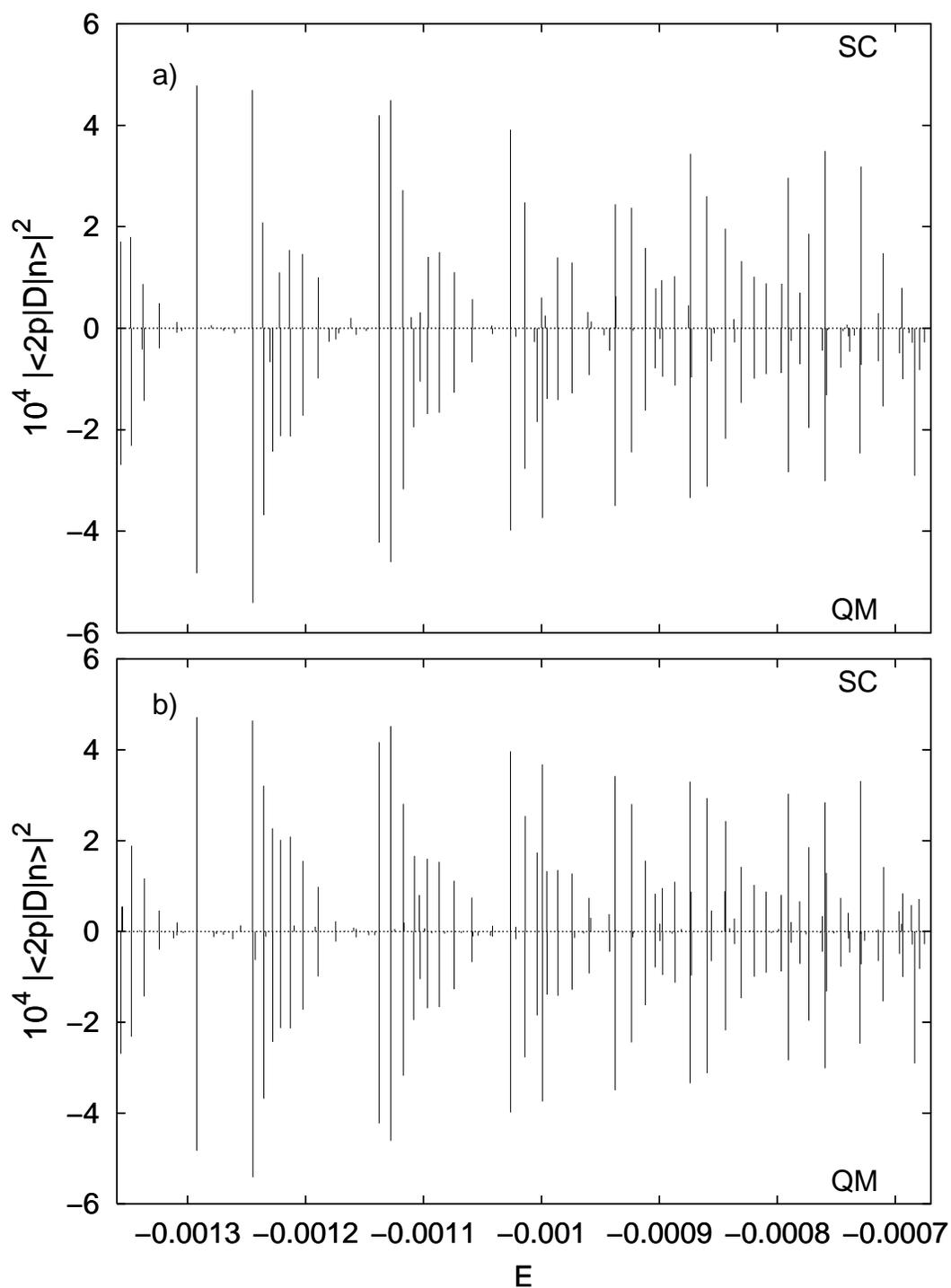}
\end{center}  
\caption{Semiclassical (SC) and quantum (QM) photoabsorption spectra
of the hydrogen atom in a magnetic field at field strength $B=11.75\,{\rm T}$ 
($\gamma = 5\times 10^{-5}\,{\rm a.u.}$).
Transitions from the initial state $|2p0\rangle$ with light polarized 
parallel to the magnetic field axis.
(a) Semiclassical spectrum obtained by harmonic inversion of a
one-dimensional recurrence signal.
(b) Semiclassical spectrum obtained by harmonic inversion of a $2\times 2$
cross-correlated recurrence signal.}
\label{fig8}
\end{figure}

Figure~\ref{fig8} presents the spectra for the photo excitation 
of the initial state $|2p0\rangle$ of the hydrogen atom in a magnetic field 
at field strength $B=11.75\,{\rm T}$ ($\gamma = 5\times 10^{-5}\,{\rm a.u.}$) 
with light polarized parallel to the magnetic field axis.
The semiclassical spectra in figure~\ref{fig8}(a) and (b) have been 
obtained by harmonic inversion of a one-dimensional recurrence signal 
and a $2\times 2$ cross-correlated signal, respectively.
In general, both semiclassical spectra are in good agreement with the 
quantum mechanical result.
A detailed comparison shows that the cross-correlation technique improves
the quality of the semiclassical spectra.
Some nearly degenerate states are not resolved with the one-dimensional
signal but are well reproduced with the cross-correlation technique.
Furthermore, the semiclassical and quantum transition matrix elements
show better agreement in figure~\ref{fig8}(b) than in 
figure~\ref{fig8}(a).
The eigenenergies and transition matrix elements of selected states
are given in table~\ref{tab1}.
\begin{table}
\caption{Selected quantum and semiclassical eigenenergies and transition 
matrix elements of the spectra shown in figure~\ref{fig8}.
The indices $s$ and $c$ refer to semiclassical data obtained by harmonic
inversion of a single (one-dimensional) and a $2\times 2$ cross-correlated
recurrence signal, respectively.}
\begin{center}
\begin{tabular}{cccccc}
$10^{3}E_n^{\rm qm}$& $10^{3}E_n^{\rm c}$ & $10^{3}E_n^{\rm s}$&
$10^{4}d_n^{\rm qm}$&$10^{4}d_n^{\rm c}$&$10^{4}d_n^{\rm s}$\\ \hline
   -1.24476 &  -1.24498 &  -1.24513 & 5.4140  &  4.6455 &  4.6925 \\
   -1.23551 &  -1.23573 &  -1.23646 & 3.6826  &  3.2088 &  2.0851\\
   -1.22796 &  -1.22811 &     --    & 2.4321  &  2.2720 &    --  \\
   -1.22126 &  -1.22130 &  -1.22222 & 2.1269  &  2.0184 &  1.1014\\
   -1.21298 &  -1.21304 &  -1.21379 & 2.1333  &  2.0866 &  1.5407\\
   -1.20237 &  -1.20228 &  -1.20263 & 1.7233  &  1.5549 &  1.4640\\
   -1.18952 &  -1.18937 &  -1.18928 & 0.9899  &  0.9813 &  1.0025
\end{tabular}
\end{center}
\label{tab1}   
\end{table}
The transition to the state at energy $E=-1.22796\times 10^{-3}$ could not 
be resolved by harmonic inversion of the single recurrence signal 
(indicated by index $s$) but is well resolved when using the cross-correlated 
signal (see values with index $c$ in table~\ref{tab1}).

\begin{figure}
\begin{center}
\includegraphics[width=.9\columnwidth]{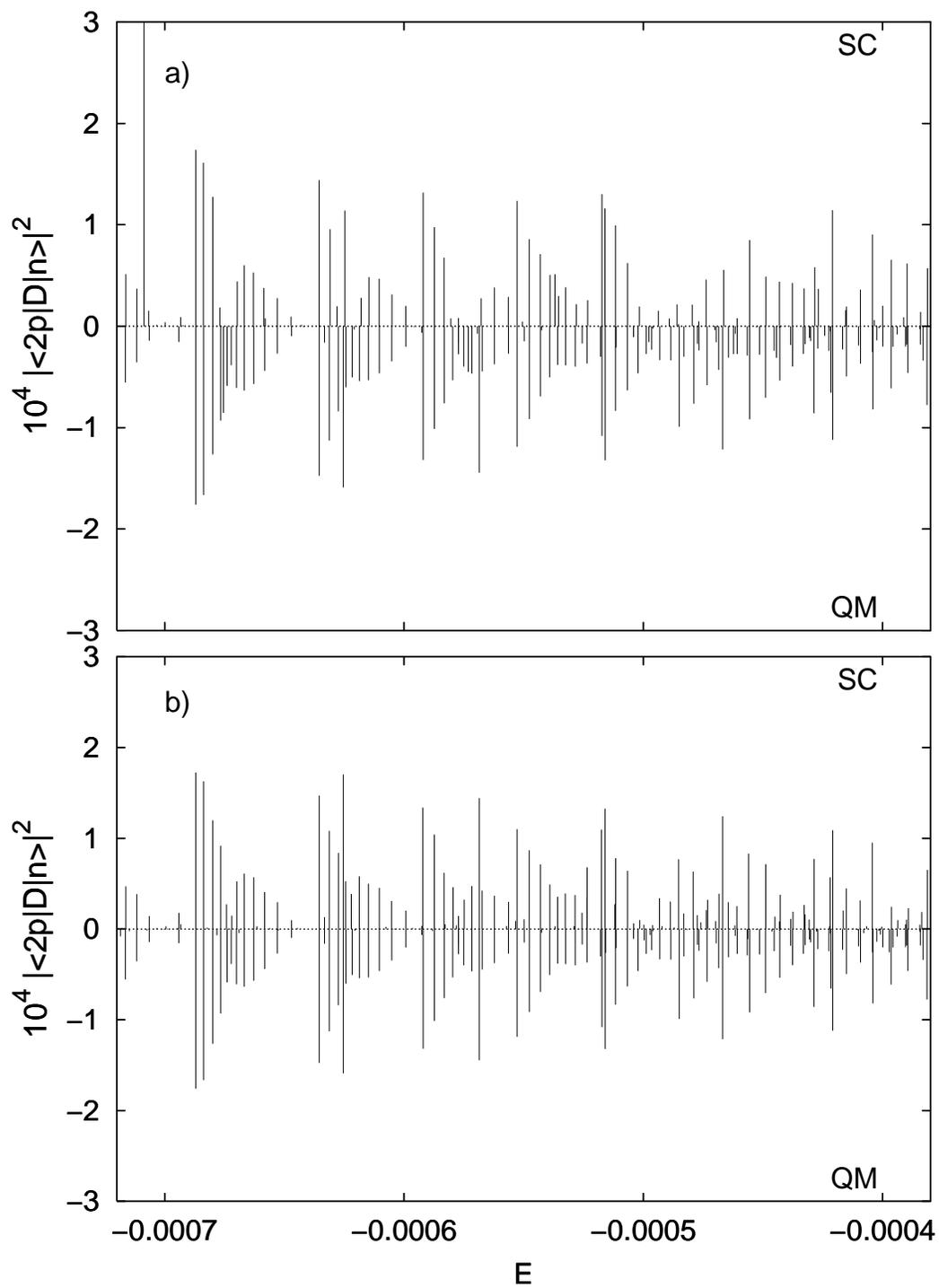}
\end{center}  
\caption{Same as figure~\ref{fig8} but for spectra at laboratory magnetic field
strength $B=4.7\,{\rm T}$ ($\gamma=2\times 10^{-5}\,{\rm a.u.}$).}
\label{fig9}
\end{figure}

In figure~\ref{fig9} the magnetic field strength is reduced to $B=4.7\,{\rm T}$
($\gamma=2\times 10^{-5}\,{\rm a.u.}$) which can be achieved easily in 
experiments.
The same closed orbit data as in figure~\ref{fig8} have been used
for the semiclassical calculations.
Similar as in figure~\ref{fig8} the semiclassical and quantum 
spectra in general are in good agreement, with the cross-correlation
technique being even more reliable than the harmonic inversion of the
one-dimensional recurrence signal.
However, the convergence of the semiclassical spectra at $B=4.7\,{\rm T}$ is
less perfect than at the higher magnetic field strength $B=11.75\,{\rm T}$.
The reason becomes evident from the scaling properties of the condition
$T_{\max}>4\pi\bar\varrho(E)$ on the required signal length.
If the scaled energy $\tilde E$ is kept constant and the magnetic field
strength $\gamma$ is varied the mean density of states scales as 
$\bar\varrho_\gamma(E=\tilde E\gamma^{2/3})
=\tilde{\bar\varrho}_{\gamma=1}(E=\tilde E)\gamma^{-4/3}$
whereas the recurrence time scales as $T=\tilde T/\gamma$.
Thus, in scaled units the required signal length reads 
$\tilde T_{\max}>4\pi\tilde{\bar\varrho}\gamma^{-1/3}$, 
and becomes larger as the field strength is decreased.
To improve the convergence properties of the harmonic inversion procedure
closed orbits with longer periods are required, i.e, the cut-off limit 
$\tilde T_{\max}$ for the scaled recurrence time must be increased.

\section{Conclusion}
\label{conclusion:sec}
Almost a century after the postulation of Bohr's quantization rules 
for the hydrogen atom and a decade and a half after the emergence 
of closed orbit theory \cite{Du88,Bog89} we have succeeded in calculating
{\em semiclassically} from first principles high-resolution photoabsorption
spectra of the diamagnetic hydrogen atom in the transition regime to chaos.
The necessary tools, viz.\ closed orbit theory, uniform approximations at
bifurcations, and the harmonic inversion method, although being known
separately, have been combined for the first time to obtain individual
semiclassical eigenenergies and transition matrix elements in that regime.

The various steps can be summarized as follows:
We have calculated all closed orbits in the energy range 
$-1\le \tilde E \le -0.5$ with recurrence times $\tilde T/2\pi\le 12$.
The rotator orbits $R_\mu^\nu$ are created in a sequence of two bifurcations,
viz.\ a pitchfork and a tangent bifurcation from the orbit perpendicular 
to the magnetic field axis.
This rather complicated scenario is described by the normal form of the 
codimension-2 symmetric butterfly catastrophe.
The vibrator orbits $V_\mu^\nu$ are created in pitchfork bifurcations of 
the parallel orbit.
Some of the symmetric vibrators and rotators undergo further pitchfork 
bifurcations, where pairs of asymmetric orbits are created.
The pitchfork bifurcations are described by the normal form of the
symmetric cusp catastrophe.
For all bifurcations of closed orbits in the selected range of energies
and recurrence times the uniform approximations have been constructed,
which remove the divergences of the isolated orbit contributions.
The contributions of the isolated closed orbits and the uniform 
approximations around bifurcations have been superimposed to obtain
semiclassical low-resolution photoabsorption spectra, and, via a 
windowed Fourier transform, the semiclassical time signal $C^{\rm sc}(t)$.
The harmonic inversion method applied to that signal finally yields the 
high-resolution spectra with individual semiclassical eigenenergies and 
transition matrix elements.
The method has been augmented by the cross-correlation technique to 
optimize its convergence properties and thus to further improve the 
quality of the results.
Spectra have been obtained at magnetic field strengths 
$B=11.75\,{\rm T}$ and $B=4.7\,{\rm T}$ in the energy region 
$-\gamma^{2/3}\le E\le -0.5\gamma^{2/3}$.
The semiclassical spectra, especially those obtained with a $2\times 2$ 
cross-correlated recurrence signal, show excellent agreement with the 
exact quantum spectra.

The semiclassical calculations can, in principle, be extended to higher
energies deep into the classically chaotic region of the diamagnetic
hydrogen atom.
However, this means that additional types of bifurcations and catastrophes
must be considered for the construction of the uniform approximations, and,
even worse, the numerical effort increases drastically due to the exponential
proliferation of closed orbits in the chaotic regime.
Clearly the objective of this paper was not to present a semiclassical method
which is computationally more efficient than exact quantum computations
(in fact, the opposite is true).
Rather, the results are of fundamental importance as regards the development, 
understanding, and practical applications of semiclassical theories.
These theories have already been successful in the limiting cases of
integrable and purely hyperbolic chaotic systems.
We have now closed the gap for systems with mixed regular-chaotic dynamics.

\ack
We thank T.~Uzer for stimulating discussions.
This work was supported in parts by the National Science Foundation and the
Deutscher Akademischer Austauschdienst.
TB is grateful to the Alexander von Humboldt-Foundation for a Feodor Lynen
fellowship.

\section*{References}
%\bibliographystyle{iopunsrt}
%\bibliography{paper}

\end{document}